# On Node Density – Outage Probability Tradeoff in Wireless Networks

Vladimir Mordachev, *Member, IEEE*, Sergey Loyka, *Senior Member, IEEE*

*Abstract*—A statistical model of interference in wireless networks is considered, which is based on the traditional propagation channel model and a Poisson model of random spatial distribution of nodes in 1-D, 2-D and 3-D spaces with both uniform and non-uniform densities. The power of nearest interferer is used as a major performance indicator, instead of a traditionally-used total interference power, since at the low outage region, they have the same statistics so that the former is an accurate approximation of the latter. This simplifies the problem significantly and allows one to develop a unified framework for the outage probability analysis, including the impacts of complete/partial interference cancelation, of different types of fading and of linear filtering, either alone or in combination with each other. When a given number of nearest interferers are completely canceled, the outage probability is shown to scale down exponentially in this number. Three different models of partial cancelation are considered and compared via their outage probabilities. The partial cancelation level required to eliminate the impact of an interferer is quantified. The effect of a broad class of fading processes (including all popular fading models) is included in the analysis in a straightforward way, which can be positive or negative depending on a particular model and propagation/system parameters. The positive effect of linear filtering (e.g. by directional antennas) is quantified via a new statistical selectivity parameter. The analysis results in formulation of a tradeoff relationship between the network density and the outage probability, which is a result of the interplay between random geometry of node locations, the propagation path loss and the distortion effects at the victim receiver.

## I. INTRODUCTION

Wireless communication networks have been recently a subject of extensive studies, both from information-theoretic and communication perspectives, including development of practical transmission strategies and fundamental limits (capacity) to assess the optimality of these strategies [1].

Mutual interference among several links (e.g. several users) operating at the same time places a fundamental limit to the network performance. The effect of interference in wireless networks at the physical layer has been studied from several perspectives [2]-[7]. A typical statistical model of interference in a network includes a model of spatial location of the nodes, a propagation path loss law (which includes the average path loss and, possibly, large-scale (shadowing) and small-scale (multipath) fading) and a threshold-based receiver performance model. The most popular choice for the model of the node spatial distribution is a Poisson point process on a plane [2]-[7]. Based on this model and ignoring the effect of fading, Sousa [3] has obtained the characteristic function (CF) of the total interference at the receiver, which can be transformed into a closed-from probability density function (PDF) in some special cases, and, based on it, the error rates for direct sequence (DS) and frequency hopping (FH) systems. For such a model, the distribution of the distance to nearest (or k-th nearest) interferer and, thus, of its interference power can be found in a compact closed form [11]-[13], [18].

While using the LePage series representation, Ilow and Hatzinakos [4][5] have developed a generic technique to obtain the CF of total interference from a Poisson point process on a plane (2-D) and in a volume (3-D), which can be used to incorporate the effects of Rayleigh and log-normal fading in a straightforward way. Relying on a homogeneous Poisson point process on a plane, Weber et al [6] have characterized the transmission capacity of the network subject to the outage probability constraint via lower and upper bounds. In a recent work, Weber et al [7] use the same approach to characterize the network transmission capacity when the receivers are able to suppress some powerful interferers, and separately include the effect of fading (based on the results in [4][5]) and of the transmission strategy [8].

A common feature of all these studies is the use of total interference, either alone or in the form of signal-to-interference-plus-noise ratio, and a common lesson is that it is very difficult to deal with: while the CF of total interference can be obtained in a closed form, the PDF or cumulative distribution function (CDF) are available only in a few special cases. This limits significantly the amount of insight that can be obtained using this approach and, thus, one has to rely on various bounds and approximations, which also complicate the analysis significantly. One notable exception is [22], where closed-form expression for the outage probability has been obtained when the required signal is subject to Rayleigh fading. This result has been extended to Nakagami-type fading in [23] and the performance of various transmission strategies has been analyzed. Unfortunately, this approach does not work when the required signal is not fading or when fading is not of Nakagami-type, or when some powerful interferers are canceled.

To overcome this difficulty, we adopt a different approach: instead of relying on the total interference power as a performance indicator, we use the power of the nearest (dominant) interferer and follow the approach originally proposed in [11]-[13]. As a result, closed-form analytical performance evaluation becomes straightforward and significant insight can be obtained using this method, including the scenarios where nearest interferers are cancelled, either via linear or nonlinear filtering techniques, and/or when interfering signals are subject to a broad class of fading processes, including all popular fading models. Further simplification by considering the low outage probability region makes the effect of various system/network parameters explicit and eliminates the need for numerical analysis of the results.

Using the methods of functions of regular variations, we show that the total interference is dominated by the nearest interferer in the region of low outage probability, i.e. the practically-important region, and, thus, both models give the same results. This result is also consistent with the corresponding results in [6]-[8], when the "near-field" region contains only one interferer. While the results in [6]-[8] hold for the uniform node density only, we consider the non-uniform case as well and also show that this conclusion holds under interference cancelation and fading.

Using this method, we study the power distribution of the dominant interferer in various scenarios, which is further used to obtain closed-form expressions for the outage probability of a given receiver or, equivalently, of the link of a given user, in the 1-D, 2-D and 3-D Poisson field of interferers, for both uniform and non-uniform node densities. Comparison to the corresponding results in [3] obtained in terms of the error rates indicates that the dominant contribution to the error rate is due to the outage events caused by the nearest interferer, which increases with the average node density. While a similar result in [6] was obtained at the low outage region in the 2-D scenario, our results hold for any outage probability and for 1, 2 and 3-D cases.

The proposed method is flexible enough to include the case when a given number of nearest interferers are canceled, either partially or completely. In the latter case, the outage probability is shown to scale down exponentially in this number. Contrary to [7], we do not rely on the simplifying assumption of canceling *all* interferers in the disk with the given average number of interferers; neither we assume that only interferers more powerful than the required signal are cancelled[1], i.e. our analysis of interference cancelation is exact. In the case of partial cancellation, we consider three different techniques and compare them using closed-form characterization of the outage probabilities, without any simplifying assumptions. The level of cancelation required to eliminate the impact of an interferer is also quantified. Proper resource allocation can significantly relax this requirement.

The proposed method is also used to include the impact of fading. Specifically, we demonstrate directly in terms of the outage probability and without using the characteristic function that the effect of a broad class of fading distributions, which includes all popular models[2] is a multiplicative constant shift of the outage probability when compared to the no-fading case. In the case of Rayleigh fading, this is a moderate constant (close to 1), and the effect of fading can be either positive (constant<1) or negative (constant>1), depending on the path loss exponent and other parameters. In the case of log-normal fading, the constant can be significantly greater than 1 and the effect of fading is always negative. The composite Rayleigh-log-normal fading results in a shift equal to the product of individual shift constants.

We further show that, for all fading distributions considered above, the total interference power is still dominated by the nearest interferer and typical outage events are due to this interferer exceeding a threshold. Thus, the outage probabilities defined in terms of the total and nearest interferer's power are the same at the low outage region. The combined effect of fading and complete/partial interference cancellation is also considered and the main conclusions above are shown to hold in this case as well. It is shown that fading relaxes the requirement to the interference cancellation level.

We observe that the outage probability versus a distortion-free interference-to-noise ratio (INR) of the receiver exhibits a threshold effect: when the distortion-free INR is below a critical value, the outage probability is high; when the distortion-free INR increases above the critical value, the outage probability sharply decreases. By quantifying the critical value, an approximation to the outage probability for the whole INR range is obtained.

Our analysis results in a formulation of the outage probability-network density tradeoff: for a given average density of the nodes, the outage probability is lower bounded or, equivalently, for a given outage probability, the average density of the nodes is upper bounded. This tradeoff is a result of the interplay between a random geometry of node locations, the propagation path loss and the distortion effects at the victim receiver.

Using the method developed, we analyze the beneficial effect of arbitrary linear filtering, e.g. by directional antennas that attenuate some interferers based on their angles of arrival, on the outage probability and on the tradeoff via a new statistical selectivity parameter (Q-parameter), which is somewhat similar to the traditional antenna gain [19] [20], but also includes the statistical distribution of interferers over the filtering variables (e.g. angles of arrival). Comparison of linear filtering to complete/partial cancellation of nearest interferers shows that the complete cancellation or partial cancellation with a sufficient cancellation level is most efficient, and that linear filtering and partial cancellation are similar in their impact on the outage probability: the latter scales linearly with the node density. Comparing our results to the corresponding linear scaling results in [23], we conclude that linear filtering has only a fixed multiplicative effect on the outage probability under a variety of scenarios, while higher-order scaling requires nonlinear interference cancelation.

---

[1] the latter assumption affects significantly the results when the threshold signal-to-interference (SIR) ratio >1, since the interferers with power below the signal power can still cause an outage but are not canceled. This explains the corresponding conclusion in [7] that the interference cancelation is only effective when the threshold SIR <1. Without such an assumption, this conclusion does not hold anymore and the interference cancellation is also effective when the threshold SIR>1 (see (19)). Thus, the results for this problem are very sensitive to the assumptions made.

[2] Rayleigh, Rice, Nakagami, log-normal, composite Rayleigh-log-normal (Suzuki), Weibull etc. [9][10]; an explicit condition for distributions to belong to this class is given.

Finally, the outage capacity is evaluated based on the results above. In particular, it is demonstrated that the effect of interference on the outage capacity is much more pronounced at low signal-to-interference ratio (SIR), and the effect of interference cancelation is much more significant at that regime as well.

The paper is organized as follows. In Section II, we introduce the system and network model. In Section III, the distribution of the interference-to-noise ratio of the nearest interferer is given for this model, including the case when most powerful interferers are cancelled. Based on this, the node density – outage probability tradeoff is presented in Section IV, including the case of complete/partial interference cancellation. The impact of fading is analyzed in Section V, the impact of linear filtering is analyzed in Section VI, and the outage capacity is evaluated in Section VII.

## II. NETWORK AND SYSTEM MODEL

As an interference model of wireless network at the physical layer, we consider a number of point-like transmitters (Tx) and receivers (Rx) that are randomly located over a certain limited region of space $S_m$, which can be one ($m = 1$), two ($m = 2$), or three ($m = 3$)-dimensional (1-D, 2-D or 3-D). This can model location of the nodes over a highway or a street canyon (1-D), a residential area (2-D), or a downtown area with a number of high-rise buildings (3-D). In our analysis, we consider a single (randomly-chosen) receiver and a number of transmitters that generate interference to this receiver. We assume that the spatial distribution of the transmitters (nodes) has the following properties: (i) for any two non-overlapping regions of space $S_a$ and $S_b$, the probability of any number of transmitters falling into $S_a$ is independent of how many transmitters fall into $S_b$, i.e. non-overlapping regions of space are statistically independent; (ii) for infinitesimally small region of space $dS$, the probability $\mathcal{P}(k = 1, dS)$ of a single transmitter ($k = 1$) falling into $dS$ is $\mathcal{P}(k = 1, dS) = \rho dS$, where $\rho$ is the average spatial density of transmitters (which can be a function of position). The probability of more than one transmitter falling into $dS$ is negligible, $\mathcal{P}(k > 1, dS) \ll \mathcal{P}(k = 1, dS)$ as $dS \to 0$. Under these assumptions, the probability of exactly $k$ transmitters falling into the region $S$ is given by Poisson distribution,

$$\mathcal{P}(k, S) = e^{-\overline{N}} \overline{N}^k / k! \qquad (1)$$

where $\overline{N} = \int_S \rho dS$ is the average number of transmitters falling into the region $S$. If the density is constant, then $\overline{N} = \rho S$. Possible scenarios to which the assumptions above apply, with a certain degree of approximation, are a sensor network with randomly-located non-cooperating sensors; a network(s) of mobile phones from the same or different providers (in the same area); a network of multi-standard wireless devices sharing the same resources (e.g. common or adjacent bands of frequencies), ad-hoc and cognitive radio networks.

Consider now a given transmitter-receiver pair. The power at the Rx antenna output $P_r$ coming from the transmitter is given by the standard link budget equation [9],

$$P_r = P_t G_t G_r g \qquad (2)$$

where $P_t$ is the Tx power, $G_t, G_r$ are the Tx and Rx antenna gains, and $g$ is the propagation path gain (=1/path loss), $g = g_a g_l g_s$, where $g_a$ is the average propagation path gain, and $g_l, g_s$ are the contributions of large-scale (shadowing) and small-scale (multipath) fading, which can be modeled as independent log-normal and Rayleigh (Rice) random variables, respectively [9].

The widely-accepted model for $g_a$ is $g_a = a_\nu R^{-\nu}$, where $\nu$ is the path loss exponent, and $a_\nu$ is constant independent of $R$ [9]. In the traditional link-budget analysis of a point-to-point link, it is a deterministic constant. However, in our network-level model $g_a$ becomes a random variable since the Tx-Rx distance $R$ is random (due to random location of the nodes) and it is this random variable that represents a new type of fading, which we term "network-scale fading", since it exhibits itself on the scale of the whole area occupied by the network. Since $g_a$ does not depend on the local propagation environment around the Tx or Rx ends that affect $g_l, g_s$ but only on the global configuration of the Tx-Rx propagation path (including the distance $R$, of which $g_l, g_s$ are independent) [9], the network-scale fading in this model is independent of the large-scale and small-scale ones, which is ultimately due to different physical mechanisms generating them[3]. Fig. 1 illustrates this. The distribution functions of $g_a$ in various scenarios have been given in [12][13].

## III. INTERFERENCE TO NOISE RATIO

We consider a fixed-position receiver (e.g. a base station of a given user) and a number of randomly located interfering transmitters (interferers, e.g. mobile units of other users) of the same power $P_t$[4]. Only the network-scale fading is taken into account in this section, assuming that $g_l = g_s = 1$ (this assumption is relaxed in section V). For simplicity, we also assume that the Tx and Rx antennas are isotropic (this assumption is relaxed in section VI), and consider the interfering signals at the receiver input.

The distribution of transmitters in space is given by (1). Transmitter $i$ produces the average power $P_{ai} = P_t g_a(R_i)$ at the receiver input. We define the interference-to-noise ratio (INR) $d_a$, also known as dynamic range [11]-[13], in the ensemble of the interfering signals via the most powerful signal at the Rx input[5],

$$d_a = P_{a1}/P_0 \qquad (3)$$

---

[3] There is a significant difference between these types of fading from the ergodicity viewpoint: while small-scale and large-scale fading require the ergodicity assumption for the statistical results to be relevant, the network-scale fading does not require it: a single instance of a network (i.e. a set of randomly located nodes) will generate appropriate empirical distribution of interference at a given receiver and thus the results of statistical analysis are applicable in this single instance (provided that the number of nodes is large enough)

[4] following the framework in [11]-[13], this can also be generalized to the case of unequal Tx powers.

[5] Theorem 1 shows that, in the small outage region, the total interference power is dominated by the contribution of the most powerful signal.

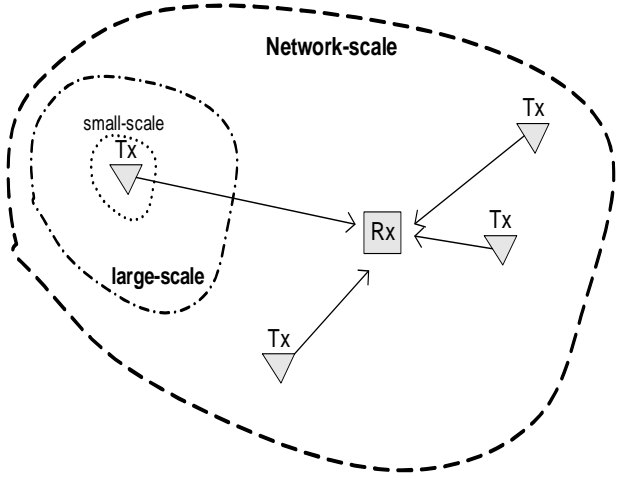

Fig. 1. Illustration of the problem geometry and three associated scales: small-scale (immediate neighborhood of a Tx; this is the scale of multipath fading), large-scale (extends beyond immediate neighborhood but is smaller than the whole network area; this is the scale of shadow fading) and network-scale (includes the whole network; this is the scale of network fading in (6)-(8)).

where $P_0$ is the noise level and, without loss of generality, we index the transmitters in the order of decreasing Rx power, $P_{a1} \geq P_{a2} \geq ... \geq P_{aN}$, and $N$ is the number of transmitters. The most powerful signal is coming from the transmitter located at the minimum distance $r_1$, $P_{a1} = P_t g_a(r_1)$. The CDF of the minimum distance can be easily found [11]-[13][18],

$$F_1(r) = 1 - \exp\left(-\overline{N}(V)\right) \quad (4)$$

where $\overline{N}(V) = \int_V \rho dV$ is the average number of transmitters in the ball $V(r)$ of radius $r$. The corresponding PDF can be found by differentiation,

$$f_1(r) = e^{-\overline{N}} \int_{V'(r)} \rho dV \quad (5)$$

where $V'(r)$ is sphere of radius $r$ and the integral in (5) is over this sphere.

The probability that the INR exceeds value $D$ is $\Pr\{d_a > D\} = \Pr\{r_1 < r(D)\} = F_1(r(D))$, where the distance $r(D)$ is such that $P_a(r(D)) = P_0 D$, so that the CDF of $d_a$ is

$$F_d(D) = 1 - \Pr\{d_a > D\} = \exp(-\overline{N}(D)) \quad (6)$$

where $\overline{N}(D) = \int_{V(r(D))} \rho dV$ is the average number of transmitters in the ball $V(r(D))$ of the radius $r(D) = (P_t a_\nu / P_0 D)^{1/\nu}$. The corresponding PDF can be obtained by differentiation,

$$f_d(D) = \frac{r(D)e^{-\overline{N}(D)}}{\nu D} \int_{V'(r(D))} \rho dV \quad (7)$$

When the average spatial density of transmitters is constant, $\rho = const$, (6), (7) simplify to [11]-[13],

$$F_d(D) = \exp\left\{-c_m \rho \left(\frac{P_t a_\nu}{P_0 D}\right)^{m/\nu}\right\} = \exp\left\{-\frac{\overline{N}_{\max}}{D^{m/\nu}}\right\},$$

$$f_d(D) = \frac{m}{\nu} \frac{\overline{N}_{\max}}{D^{m/\nu+1}} \exp\left\{-\frac{\overline{N}_{\max}}{D^{m/\nu}}\right\} \quad (8)$$

where $c_1 = 2$, $c_2 = \pi$ and $c_3 = 4\pi/3$, $\overline{N}_{\max} = c_m R_{\max}^m \rho$ is the average number of transmitters in the ball of radius $R_{\max}$, which we term "potential interference zone", and $R_{\max} = r(1) = (P_t a_\nu / P_0)^{1/\nu}$ is such that $P_a(R_{\max}) = P_0$, i.e. a transmitter at the boundary of the potential interference zone produces signal at the receiver exactly at the noise level; transmitters located outside of this zone produce weaker signals, which are neglected in the interference-limited scenario (see Fig. 2). Note that (8) gives the distribution of the INR as an explicit function of the system and geometrical parameters, and ultimately depends on $\overline{N}_{\max}, m, \nu$ only.

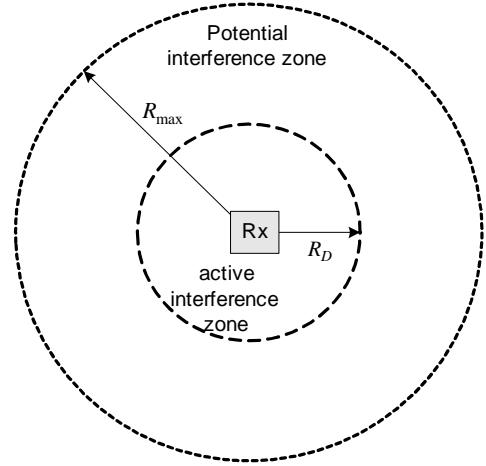

Fig. 2. Interference zones on the network scale. Potential interference zone: $R \leq R_{\max}, P_a(R) \geq P_0 = P_a(R_{\max})$, i.e. the interference power exceeds the Rx noise level; when evaluating the total interference power, only the interferers in this zone are considered (this corresponds to either the fact that radio waves at GHz frequencies decay exponentially fast when blocked by earth curvature or the area populated by interferers being finite). Active interference zone: $R \leq R_D, P_a(R) \geq P_{df} = P_a(R_D)$, i.e. the interference power exceeds the maximum distortion-free level.

When $(k-1)$ most powerful signals, which are coming from $(k-1)$ closest transmitters, do not create any interference (i.e. due to frequency, time or code separation in the multiple access scheme, or due to any other form of separation or filtering), the CDF and PDF of the distance $r_k$ to the most powerful interfering signal of order $k$ can be found in a similar way. The CDF of the INR $d_a$ in this case is given by

$$F_{dk}(D) = e^{-\overline{N}(D)} \sum_{i=0}^{k-1} \overline{N}(D)^i / i! \quad (9)$$

In the case of constant average density $\rho = const$, the CDF and PDF of the INR simplify to [11]-[13],

$$F_{dk}(D) = \exp\left\{-\frac{\overline{N}_{\max}}{D^{m/\nu}}\right\} \sum_{i=0}^{k-1} \frac{1}{i!} \left(\frac{\overline{N}_{\max}}{D^{m/\nu}}\right)^i,$$

$$f_{dk}(D) = \frac{m}{\nu(k-1)!} \frac{\overline{N}_{\max}^k}{D^{\frac{km}{\nu}+1}} \exp\left\{-\frac{\overline{N}_{\max}}{D^{m/\nu}}\right\} \quad (10)$$

## IV. OUTAGE PROBABILITY-NODE DENSITY TRADEOFF

Powerful interfering signals can result in significant performance degradation due to linear and nonlinear distortion effects in the receiver when they exceed certain limit, which we characterize here via the receiver distortion-free INR, i.e. the maximum acceptable interference-to-nose ratio, $D_{df} = P_{\max}/P_0$, where $P_{\max}$ is the maximum interference power at the receiver that does not cause significant performance degradation. This is equivalent to using the signal to noise plus interference ratio when the required signal power is fixed, e.g. no or negligible fading due to strong line of sight component. If $d_a > D_{df}$, there is significant performance degradation and the receiver is considered to be in outage, which corresponds to one or more transmitters falling into the active interference zone (i.e. the ball of radius $r(D_{df})$; the signal power coming from transmitters at that zone exceeds $P_{\max}$, whose probability is

$$\mathcal{P}_{out} = \Pr\{d_a > D_{df}\} = 1 - F_d(D_{df}) \quad (11)$$

For given $\mathcal{P}_{out}$, one can find the required distortion-free INR ("outage INR") $D_{df}$

$$D_{df} = F_d^{-1}(1 - \mathcal{P}_{out}) \quad (12)$$

We note that, in general, $D_{df}$ is a decreasing function of $\mathcal{P}_{out}$, i.e. low outage probability calls for high distortion-free INR. For simplicity of notations, we further drop the subscript and denote the distortion-free INR by $D$.

While the definition of outage probability above relies on the maximum interfering power, the same outage probability holds in terms of the total interfering power at the low outage region, as the theorem below demonstrates.

*Theorem 1:* Consider the outage probability in (11). At the low outage region, it converges to the outage probability defined via the total interference power, i.e.

$$\lim_{x \to \infty} \frac{\Pr\{\sum_i P_{ai} > x\}}{\Pr\{P_{a1} > x\}} = 1 \quad (13)$$

*Proof:* via the functions of regular variation; see Appendix 1 for details.

Thus, at the low outage region, $\mathcal{P}_{out}$ in (11) serves as an accurate approximation of the outage probability in terms of the total interference power,

$$\Pr\left\{\sum_i P_{ai} > x\right\} \approx \Pr\{P_{a1} > x\}, \text{ for large } x, \quad (14)$$

and all our results also apply to such an outage probability. A significant advantage of (11) is that a closed-form analysis becomes straightforward.

### A. All interfering signals are active ($k = 1$)

We consider first the case of $k = 1$, i.e. all interfering signals are active. The outage probability can be evaluated using (6) and (11). From practical perspective, we are interested in the range of small outage probabilities $\mathcal{P}_{out} \ll 1$, i.e. high-reliability communications. When this is the case, $F_d(D) \to 1$ and using MacLaurean series expansion $e^{-\overline{N}} \approx 1 - \overline{N}$, where $\overline{N}$ is the average number of transmitters in the active interference zone, (11) simplifies to

$$\mathcal{P}_{out} \approx \overline{N} = \int_{V(r(D))} \rho dV \quad (15)$$

which further simplifies, in the case of $\rho = const$, to

$$\mathcal{P}_{out} \approx \overline{N}_{\max} D^{-m/\nu} \quad (16)$$

Note that, in this case, the outage probability $\mathcal{P}_{out}$ scales linearly with the average number $\overline{N}_{\max}$ of nodes in the potential interference zone and also with the node density $\rho$, and it effectively behaves as if the number of nodes were fixed (not random) and equal to $\overline{N}_{\max}$. Based on this, we conclude that the single-order events (i.e. when only one signal in the ensemble of interfering signals exceeds the threshold $P_{\max}$) are dominant contributor to the outage, which is also consistent with Theorem 1. This immediately suggests a way to reduce significantly the outage probability by eliminating the dominant interferer in the ensemble. Using (16), the required spurious-free INR of the receiver can be found for given outage probability, $D \approx (\overline{N}_{\max}/\mathcal{P}_{out})^{\nu/m}$. Note that higher values of $\nu$ and lower values for $m$ call for higher distortion-free INR. Intuitively, this can be explained by the fact that when the transmitter moves from the boundary of the potential interference zone (i.e. $R = R_{\max}$, $P_a(R) = P_0$) closer to the receiver ($R \ll R_{\max}$), the power grows much faster when $\nu$ is large, so that closely-located transmitters produce much more interference (compared to those located close to the boundary) in that case, which, combined with the uniform spatial density of the transmitters, explains the observed behavior. The effect of $m$ can be explained in a similar way.

To validate the accuracy of approximation in (15), and also the expressions for the INR's PDF and CDF in the previous section, extensive Monte-Carlo (MC) simulations have been undertaken. Fig. 3 shows some of the representative results. Note good agreement between the analytical results (including the approximations) and the MC simulations. It can be also observed that the tails of the distributions decay much slower for the $\nu = 4$ case, which indicates higher probability of high-power interference in that case and, consequently, requires higher distortion-free INR of the receiver, in complete agreement with the predictions of the analysis. Note also that the outage probability evaluated via the total interference power coincides with that evaluated via the maximum interferer power (at the small outage region), in complete agreement with Theorem 1.

Consider now a scenario where the actual outage probability should not exceed a given value $\epsilon$, $\mathcal{P}_{out} \leq \epsilon$, for the receiver with a given distortion-free INR $D$. Using (8) and (11), the average number of transmitters in the active interference zone (ball of radius $r(D)$) can be upper bounded as $\overline{N} \leq -\ln(1 - \epsilon)$. Using the expression for $\overline{N}$, one obtains a basic tradeoff relationship between the network density and the outage probability,

$$\overline{N} = \int_{V(r(D))} \rho dV \leq -\ln(1 - \epsilon) \approx \epsilon \quad (17)$$

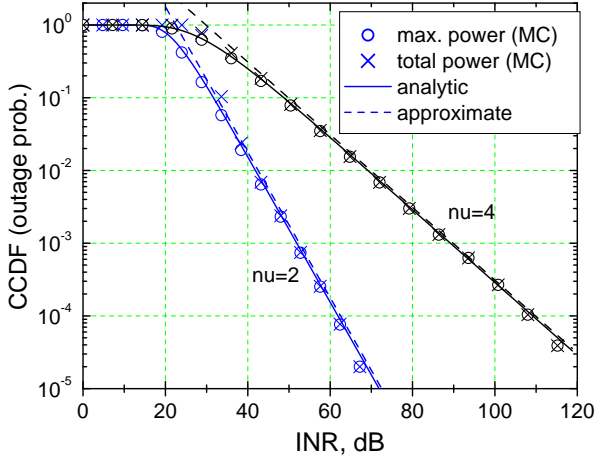

Fig. 3. The CCDF of $d_a = P_{a1}/P_0$ and $d_{tot} = P_{tot}/P_0$ (also the outage probability) evaluated from Monte-Carlo (MC) simulations for $m = 2$, $\nu = 2\&4$, $P_0 = 10^{-10}$, $P_t = 1$, $\rho = 10^{-5}$; analytic CCDF of $d_a$ (derived from (8)) and its approximation in (16) are also shown. Note that the approximation becomes very accurate at $\mathcal{P}_{out} \leq 0.1$ and that the CCDF of total and maximum interference power are the same at this region.

where the approximation holds in the small outage region. Thus, for given outage probability, the network density is upper bounded or, equivalently, for given network density, the outage probability is lower bounded.

In the case of uniform density $\rho = const$ and small outage probability, $\epsilon \ll 1$, this gives an explicit tradeoff relationship between the maximum distortion-free interference power at the receiver $P_{\max}$, the transmitter power $P_t$ and the average node density for distortion-free receiver operation,

$$\rho \leq c_m^{-1} \epsilon \left( P_{\max}/P_t a_\nu \right)^{m/\nu} \qquad (18)$$

or, equivalently, an upper bound on the average density of nodes in the network. As intuitively expected, higher $\epsilon, P_{\max}, \nu$ and lower $P_t, m$ allow for higher network density. The effect of $\nu$ is intuitively explained by the fact that higher $\nu$ results in larger path loss or, equivalently, in smaller distance at the same path loss, so that the transmitters can be located more densely without significant increase in the interference level. The effect of the other parameters can be explained in a similar way.

### B. $(k-1)$ nearest interferers are cancelled

We now assume that $(k-1)$ nearest interferers are eliminated via some means (e.g. by processing at the receiver or resource allocation). In this case, (9), (10) apply and (15) generalizes to

$$\mathcal{P}_{out} \approx \frac{1}{k!} \overline{N}^k = \frac{1}{k!} \left( \frac{\overline{N}_{\max}}{D^{m/\nu}} \right)^k \qquad (19)$$

which can be expressed as $\mathcal{P}_{out} = \frac{1}{k!} \mathcal{P}_{out,1}^k \leq \mathcal{P}_{out,1}$, where $\mathcal{P}_{out,1}$ is the outage probability for $k = 1$ (see (15)). In the small outage region, $\mathcal{P}_{out,1} \ll 1$ and $\mathcal{P}_{out} \ll \mathcal{P}_{out,1}$, i.e. there is a significant beneficial effect of removing $(k-1)$ strongest interferers, which scales exponentially with $k$. Fig.

4 illustrates this case. Note that the outage probabilities in terms of maximum and total interference power are close to each other at the low outage region.

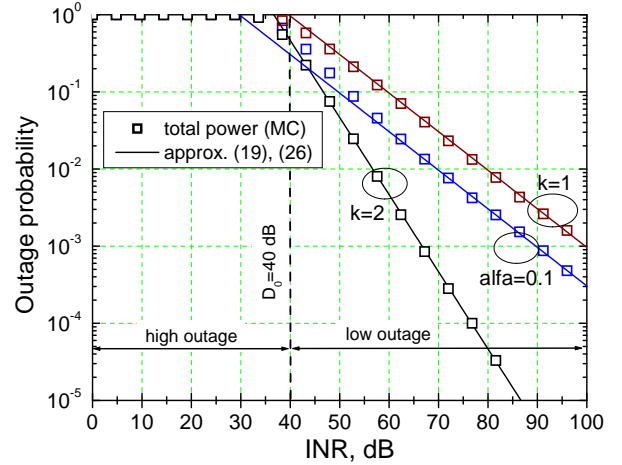

Fig. 4. The outage probability vs. INR for $k = 1$ (no cancelation), $k = 2$ (nearest interferer is canceled) and $\alpha = 0.1$ (partial cancelation) via the total power (MC) and the approximations in (19), (26); $\nu = 4$, $m = 2$, $\overline{N}_{max} = 100$, $R_{max} = 10^3$. While the full cancelation results in exponential scaling of $\mathcal{P}_{out}$ with $k$, partial cancelation results in no exponential scaling but only fixed improvement of 10 dB. The threshold effect is clear and the critical INR is $D_0 \approx 40dB$. The outage probability via the total power agrees well with the approximations.

Further comparison to the corresponding result in [7], which was obtained under the assumption of cancelling *all* interferers that exceed the required signal and are in the disk with the given average number of interferers, shows that this assumption affects significantly the outage probability, resulting in no exponential scaling and ultimately responsible for the conclusion in [7] that interference cancellation is effective only when the threshold SIR $< 1$. If this assumption is removed, the interference cancellation is effective for any SIR, as (19) demonstrates.

From the MacLaurean series expansion of $F_{dk}(D)$ in (9), (10) in $1/D$, the approximation in (19) becomes accurate when $\overline{N} < 1$, i.e. for the uniform density,

$$D > \overline{N}_{\max}^{\nu/m} \qquad (20)$$

It should also be noted that $\mathcal{P}_{out}$ in (19) scales as $\overline{N}_{\max}^k$ or as $\rho^k$, i.e. it is much more sensitive to the node density in this case, and the sensitivity increases exponentially with $k$.

Fig. 3 - 5 reveal a threshold effect: when the distortion-free INR is below a critical value $D_0$, the outage probability is high, $\mathcal{P}_{out} \approx 1$; for $D > D_0$, $\mathcal{P}_{out}$ sharply decreases and the approximation in (19) becomes accurate, so that the distortion-free INR should be higher than $D_0$ to keep $\mathcal{P}_{out}$ low. When $k$ is not too large, the critical INR corresponds to on average one interferer being in the active interference zone $\overline{N}(D_0) = 1$, since this causes high $\mathcal{P}_{out}$ [6], so that for the uniform density of interferers,

$$D_0 \approx \overline{N}_{\max}^{\nu/m} \qquad (21)$$

---

[6]When $\overline{N} = 1$, $\mathcal{P}_{out} = 1 - e^{-1} \approx 0.63$ for $k = 1$, and $\mathcal{P}_{out} = 1 - 2e^{-1} \approx 0.26$ for $k = 2$, i.e. $\mathcal{P}_{out}$ is high unless $k \gg \overline{N}$.

i.e. the critical INR is directly related to the average number of interferers in the potential interference zone. Based on this threshold effect, we propose a piece-wise linear (on log-log scale) approximation of $\mathcal{P}_{out}$ for the whole INR range,

$$\mathcal{P}_{out} \approx \begin{cases} 1, & D < D_0 \\ \text{eq. 19}, & D > D_0 \end{cases} \quad (22)$$

In a similar way, the node density-outage probability trade-off can be formulated. In the for small outage probability region $\epsilon \ll 1$, it can be expressed as

$$\overline{N} = \int_{V(r(D))} \rho dV \leq (k!\epsilon)^{1/k} \quad (23)$$

Comparing (23) to (17), one can clearly see the beneficial effect of "removing" $(k-1)$ most powerful interferers on the outage probability-network density tradeoff, since $(k!\epsilon)^{1/k} \gg \epsilon$ in the small outage regime, so that higher node density is allowed at the same outage probability.

In the case of uniform density, (23) reduces to

$$\rho \leq c_m^{-1} (k!\epsilon)^{1/k} (P_{\max}/P_t a_\nu)^{m/\nu} \quad (24)$$

which is a generalization of (18) to $k > 1$. Note that the upper bound on the node density scales as $(k!\epsilon)^{1/k}$, i.e. much better than in (18).

### C. Partial cancellation of $(k-1)$ nearest interferers

Following [7], one can also consider the case of non-ideal (realistic) interference cancellation, when $(k-1)$ nearest interferers are attenuated by a factor of $0 \leq \alpha \leq 1$ (so that the interference power is $\alpha P_{ai}$, $1 \leq i \leq k-1$) where $\alpha = 0$ corresponds to the ideal case (complete cancellation) and $\alpha = 1$ corresponds to the case of no cancellation at all. When $\alpha$ is independent of $D$, it is straightforward to show that asymptotically ($D \to \infty$) the nearest interferer dominates the outage probability, which is given by

$$\mathcal{P}_{out} \approx \alpha^{m/\nu} \overline{N}_{\max} D^{-m/\nu}, \ \alpha > 0 \quad (25)$$

and which is also the same as that of partially cancelling only the nearest interferer ($k = 2$), i.e. partial cancelling of more than one nearest interferer by a fixed level does not bring any additional advantage asymptotically, and the outage probability in this case significantly exceeds that of complete cancellation (compare (25) to (19)). Comparing (25) to (16), the effect of partial cancellation by a factor of $\alpha$ is to reduce $\mathcal{P}_{out}$ by a factor of $\alpha^{m/\nu}$ compared to the no cancellation case, i.e. by a factor of $\alpha$ for $m = 2$ and $\nu = 2$ (free space propagation) and by a factor of $\sqrt{\alpha}$ for $m = 2$ and $\nu = 4$ (two-ray propagation or ground reflection). Fig. 4 illustrates this case.

One can also consider another scenario, where $(k-2)$ nearest interferers are cancelled completely (for example, by proper resource allocation, frequency or time) and $(k-1)$-th interferer is cancelled partially (e.g. by processing at the receiver), $k \geq 3$. In such a case, it is straightforward to show that the $(k-1)$-th interferer dominates asymptotically and the outage probability is given by

$$\mathcal{P}_{out} \approx \frac{\alpha^{(k-1)m/\nu}}{(k-1)!} \left(\frac{\overline{N}_{\max}}{D^{m/\nu}}\right)^{k-1}, \ \alpha > 0 \quad (26)$$

i.e. a significant improvement over (25), but still higher than (19) (complete cancellation). Similarly to (20), the approximation in (26) becomes tight when

$$D > \overline{N}_{\max}^{\nu/m}/\alpha^{k-1} \quad (27)$$

For the whole distortion-free INR range, one can use (22) in combinaion with (26).

Finally, one can also consider the case where $\alpha$ scales as a function of $D$ and ask a question: "*What level of cancellation is required to eliminate the effect of $(k-1)$-th nearest interferer?*" Assuming that $(k-2)$ nearest interferers are cancelled completely and comparing the contribution of the partially-cancelled $(k-1)$-th interferer (see (26)) to $k$-th interferer (not cancelled at all, see (19)), it is straightforward to show that the $k$-th interferer dominates if

$$\alpha < \frac{1}{D^{1/(k-1)}} \left(\frac{\overline{N}_{\max}}{k}\right)^{\nu/m(k-1)} \quad (28)$$

Thus, perfect cancellation is not a prerequisite and $\alpha > 0$ can also do the job, if it properly scales with $D$.

Similar condition can also be obtained when $(k-1)$ nearest interferers are partially cancelled by the same factor $\alpha$,

$$\alpha < \frac{1}{D^{k-1}} \left(\frac{\overline{N}_{\max}^{k-1}}{k!}\right)^{\nu/m} \quad (29)$$

which is, however, a significantly tighter requirement than (28), as intuitively expected. Thus, complete cancellation of some nearest interferers (e.g. via resource allocation) is of significant help when only partial (realistic) cancellation at the receiver is possible.

### D. Total Interference Power

If the total interference power is used to define the outage probability, the results will be the same in the small outage region, as indicated by the following theorem (equivalent of Theorem 1).

*Theorem 2:* Consider the outage probability in (19). At the low outage region, it converges to the outage probability defined via the total interference power, i.e.

$$\lim_{x \to \infty} \frac{\Pr\left\{\sum_{i=k}^{N} P_{ai} > x\right\}}{\Pr\{P_{ak} > x\}} = 1 \quad (30)$$

and, thus, the following approximation holds,

$$\Pr\left\{\sum_{i=k}^{N} P_{ai} > x\right\} \approx \Pr\{P_{ak} > x\}, \text{ for large } x. \quad (31)$$

*Proof:* along the same lines as that of Theorem 1.

Fig. 4 validates this Theorem via Monte-Carlo simulations. We also note that this Theorem also applies when a partial interference cancellation (as above) is considered and, thus, the outage probabilities in (25), (26) also hold in terms of the total interference power.

## V. IMPACT OF FADING

In this section, we study the impact of fading directly in terms of the outage probability, which provides additional insight into interference-generating mechanisms and their impact. In particular, we demonstrate that the total interference power is dominated by that of the nearest interferer for a broad class of fading distributions, including all popular models. This also holds when some nearest interferers are canceled.

### A. Impact of Rayleigh fading

Let us consider the ordered average powers $P_{a1} \geq P_{a2} \geq ... \geq P_{aN}$ which are further subjected to Rayleigh fading so that the fading received powers are $P_{si} = g_{si} P_{ai}$, where $g_{si}$ are the Rayleigh fading factors, assumed to be i.i.d., with the standard pdf $f_{gs}(x) = e^{-x}$. The INR is now defined as $d_s = P_{s1}/P_0 = d_a g_{s1}$, where $d_a = P_{a1}/P_0$, i.e. via the contribution of the nearest interferer[7], and its cumulative CDF (CCDF), i.e. the outage probability, is

$$\mathcal{P}_{out} = \Pr\{d_s > D\} = \int_0^\infty f_{gs}(g) \overline{F}_d(D/g) \, dg, \quad (32)$$

where $\overline{F}_d(x) = 1 - F_d(x)$ is the CCDF of $d_a$. At the low outage region, i.e. at the distribution tail $D \to \infty$, it can be approximated as

$$\mathcal{P}_{out} \approx \Gamma(m/\nu + 1) \overline{N}_{\max} D^{-m/\nu}, \quad (33)$$

where $\Gamma$ is the gamma function (see Appendix 2 for proof). Comparing to (16), we conclude that the effect of Rayleigh fading is the multiplicative shift by a constant factor $\Gamma(m/\nu + 1)$, and the functional form of the distribution (i.e. regular variation or heavy tail) is preserved. Since $\Gamma(m/\nu + 1)$ can be greater or smaller than 1, depending on $m/\nu$ (e.g. $m = 2, \nu = 4 \to \Gamma \approx 0.89$), the effect of Rayleigh fading can be both positive and negative.

In a similar way, one obtains the outage probability when $(k-1)$ nearest interferers are cancelled,

$$\mathcal{P}_{out} \approx \frac{\Gamma(km/\nu + 1)}{k!} \left(\frac{\overline{N}_{\max}}{D^{m/\nu}}\right)^k, \quad (34)$$

i.e. the beneficial effect of cancelling is slightly offset by fading (since $\Gamma(km/\nu + 1)$ is an increasing function of $k$) but otherwise follows the same tendency as without fading (see (19)).

Since the INR is the scaled interference power, the later will follow the same distribution as in (34) (up to a constant) and, thus, $\Pr\{P_{sk} > x\}$ is a function of regular variation so that Theorem 2 applies, i.e.

_Theorem 3:_ When the interferers are subject to the average path loss and Rayleigh fading, the nearest interferer dominates in terms of the outage probability at the low outage region, i.e.

$$\lim_{x \to \infty} \frac{\Pr\left\{\sum_{i=k}^{N} P_{si} > x\right\}}{\Pr\{P_{sk} > x\}} = 1 \quad (35)$$

[7] which may sometimes be not the largest one (due to the effect of Rayleigh fading). However, as we show below, the nearest interferer contribution dominates the tail of the total interference distribution and thus the outage probability.

and,

$$\Pr\left\{\sum_{i=k}^{N} P_{si} > x\right\} \approx \Pr\{P_{sk} > x\}, \text{ for large } x \quad (36)$$

Thus, the results in (33), (34) also apply to the outage probability defined via the total interference power. This complements the results in [4] obtained in terms of the characteristic function with compact, closed-form expressions for the outage probability and also explicitly demonstrates the effect of cancelling $(k-1)$ nearest interferers. Fig. 5 illustrates this case.

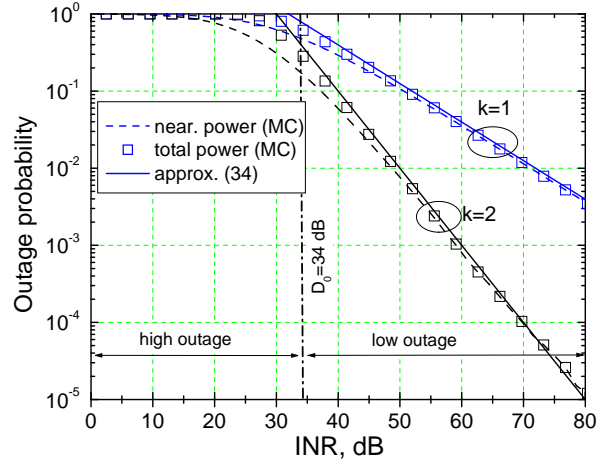

Fig. 5. The outage probability vs. INR for $k = 1$ (no cancellation) and $k = 2$ (nearest interferer is canceled) via nearest and total power under Rayleigh fading; $\nu = 4$, $m = 2$, $\overline{N}_{max} = 50$, $R_{max} = 10^3$. The exponential scaling of $\mathcal{P}_{out}$ with $k$ is preserved under fading, as well as the dominance of the nearest interferer in the small outage region. The threshold effect is clear and the critical INR is $D_0 = 34$ dB.

The intuition behind Theorem 3 is that the distributions in (11), (16), (19) are much more heavily-tailed (slowly-decaying) than the Rayleigh distribution so that outage events in the combined distribution are mostly caused by nearby interferers without deep Rayleigh fades and the combined distribution is a slightly shifted version of the original one (without fading).

### B. Impact of log-normal and combined fading

This can be analysed in a similar way. The main results are summarized as follows. When the interferers are subject to the average path loss and log-normal i.i.d. fading, and when $(k-1)$ nearest interferers are cancelled, the outage probability is

$$\mathcal{P}_{out} \approx \frac{M_{km/\nu}}{k!} \left(\frac{\overline{N}_{\max}}{D^{m/\nu}}\right)^k, \quad (37)$$

where $M_{km/\nu} = \exp\left(\frac{1}{2}(\sigma km/\nu)^2\right)$ is $km/\nu$-th moment of the log-normal random variable,

$$M_{km/\nu} = \frac{1}{\sqrt{2\pi}\sigma} \int_0^\infty x^{km/\nu - 1} \exp\left(-\frac{(\ln x)^2}{2\sigma^2}\right) dx, \quad (38)$$

and $\sigma$ is the standard deviation. The case when no interferers are cancelled corresponds to $k = 1$. Comparing (37) to (19), we conclude that the effect of log-normal fading is a shift by a constant factor $> 1$, i.e. strictly negative as opposed to Rayleigh fading where it can be either positive or negative. The beneficial effect of cancelling $(k-1)$ nearest interferers is also offset by fading in this case (since $M_{km/\nu}$ increases with $k$). Since the regular varying (heavy tail) nature of the distribution is preserved, Theorem 3 also holds in this case, i.e. the nearest interferer is still dominant.

Likewise, one can consider the combined effect of Rayleigh and log-normal fading. The outage probability is

$$\mathcal{P}_{out} \approx \frac{\Gamma(km/\nu+1)M_{km/\nu}}{k!}\left(\frac{\overline{N}_{\max}}{D^{m/\nu}}\right)^k, \quad (39)$$

and Theorem 3 also applies. Note that the effects of Rayleigh and log-normal fading are multiplicative in terms of the shift constant, and the heavy tail of the distribution, which is due to the Poisson spatial distribution of the interferers and the average path loss, is not affected.

### C. The impact of a broad class of fading distributions

The results above are not limited to Rayleigh or log-normal fading but rather hold for a broad class of distributions whose tails are dominated by the tail of $P_{ak}$.

*Theorem 4:* Let the interferers be subject to the average path loss and fading, $P_i = g_i P_{ai}$, where $g_i$ is the fading power gain, i.i.d. for each interferer, and the fading distribution tail is dominated by that in (19), i.e.,

$$\lim_{x \to \infty} \Pr\left(g_i > x\right) x^{km/\nu} = 0, \quad (40)$$

then the outage probability is

$$\mathcal{P}_{out} \approx \frac{M_{km/\nu}}{k!}\left(\frac{\overline{N}_{\max}}{D^{m/\nu}}\right)^k, \text{ for large } D, \quad (41)$$

where $M_{km/\nu}$ is $km/\nu$-th moment of the fading power gain, $M_{km/\nu} = \int_0^\infty x^{km/\nu} f_g(x) dx$, and $f_g(x)$ is the pdf of $g$. Furthermore, the nearest interferer dominates the outage events, i.e. Theorem 3 holds, and, thus, the outage probability in (39) holds in terms of both the total and nearest interferer's power.

*Proof:* along the same lines as that of Theorem 3 and that of (33).

It should be noted that Theorem 4 includes almost all popular fading models, i.e. Rayleigh, Rice, Nakagami, Weibul, log-normal, or any distribution whose tail decays faster than polynomially. It is also interesting to note that the fading enters the outage probability only via the moment $M_{km/\nu}$ and the condition of tail dominance, all other details being irrelevant. The effect of fading is positive for $M_{km/\nu} < 1$ and negative for $M_{km/\nu} > 1$.

Finally, the effect of fading can also be considered jointly with partial interference cancellation, and the outage probabilities in (25), (26) are respectively modified to

$$\mathcal{P}_{out} \approx \alpha^{m/\nu} M_{m/\nu} \overline{N}_{\max} D^{-m/\nu} \quad (42)$$

$$\mathcal{P}_{out} \approx \frac{\alpha^{(k-1)m/\nu} M_{(k-1)m/\nu}}{(k-1)!}\left(\frac{\overline{N}_{\max}}{D^{m/\nu}}\right)^{k-1} \quad (43)$$

i.e. the multiplicative constant shift of the outage probability is preserved. The required partial cancellation levels in (28) and (29) are modified to

$$\alpha < \frac{1}{D^{1/(k-1)}}\left(\frac{M_{km/\nu}\overline{N}_{\max}}{M_{(k-1)m/\nu} \cdot k}\right)^{\nu/m(k-1)} \quad (44)$$

$$\alpha < \frac{1}{D^{k-1}}\left(\frac{M_{km/\nu}\overline{N}_{\max}^{k-1}}{M_{m/\nu} \cdot k!}\right)^{\nu/m} \quad (45)$$

Noting that since $M_{km/\nu}$ increases with $k$ for Rayleigh, log-normal and composite fading, its effect on the required cancellation level is beneficial in both cases (i.e. higher $\alpha$ is acceptable), and it is more pronounced for the case of partial cancellation of $(k-1)$ nearest interferers. This is intuitively explained by the fact that these fading distributions decay very fast (exponentially or sub-exponentially) at the large signal region but only polynomially at the low signal region and, thus, the fading results more often in a weaker signal than in a stronger one.

## VI. THE IMPACT OF LINEAR FILTERING

In the previous sections, we considered the interfering signals at the Rx input assuming that the Rx antenna was isotropic, i.e. no measures to eliminate some of the interfering signals by linear filtering at the receiver were considered. In this section, we explore the effect of linear filtering, which may include filtering by the Rx antenna based on the angle of arrival, polarization and frequency, and by linear frequency filters at the receiver (at RF, IF and possibly basedband). Since, as it follows from the previous section, the average number of interfering signals $\overline{N}$ is a key parameter, which determines the INR of interfering signals (see (6),(9)) and ultimately the network density-outage probability tradeoff (e.g. (17), (23)), we consider the impact of linear filtering on this parameter. For simplicity, we further assume no nearest interference cancellation and no fading. The impact of these factors can be incorporated into the analysis in a straightforward way following the results in sections IV and V. We also assume for simplicity that the node density is uniform.

Let $\mathbf{z} = [z_1, z_2...z_l]^T$ be the set of filtering variables (i.e. frequency, polarization, angle of arrival etc.) and $f_z(\mathbf{z})$ be the PDF of incoming interfering signals over these variables. The probability of a randomly-chosen input signal (arriving from a randomly-selected node) falling in the interval $d\mathbf{z}$ is $f_z(\mathbf{z})d\mathbf{z}$, and the probability that the filter output power of this signal exceeds the threshold $P_0$ is

$$\Pr\{P_{a,out} > P_0\} = \int_{P_0/K(\mathbf{z})}^{\infty} w_a(P)dP = K^{m/\nu}(\mathbf{z}) \quad (46)$$

where $0 \le K(\mathbf{z}) \le 1$ is the normalized filter power gain (e.g. antenna pattern), and $w_a(P) = \frac{m}{\nu}P_0^{m/\nu}P^{-1-m/\nu}$, $P \ge P_0$, is the PDF of the signal power $P$. Note that $K^{m/\nu}$ represents the reduction in probability of signal power exceeding the threshold $P_0$ from the input (where it is equal to one) to the output of the filter and thus is a filter gain for given values of filtering variables. The average number of output signals exceeding the threshold in the interval $d\mathbf{z}$ is $d\overline{N}_{out} = K^{m/\nu}(\mathbf{z})f_z(\mathbf{z})d\mathbf{z}d\overline{N}_{in}$, where $d\overline{N}_{in}$ is the average number of input signals exceeding the threshold in the same interval. Finally, the total average number of output signals exceeding the threshold $P_0$ is

$$\overline{N}_{out} = \overline{N}_{in}/Q, \quad Q = \left(\int_{\Delta \mathbf{z}} K^{m/\nu}(\mathbf{z})f_z(\mathbf{z})d\mathbf{z}\right)^{-1} \geq 1 \quad (47)$$

where $\overline{N}_{in}$ is the average number of input signals, $Q$ is the average statistical filter gain, which represents its ability to reduce the average number of visible (i.e. exceeding the threshold) interfering signals, and $\Delta \mathbf{z}$ is the range of filtering variables. This gain further transforms into reduction in the INR (see (6), (9)) or the outage probability,

$$\mathcal{P}_{out} = 1 - e^{-\overline{N}_{out}} \approx \overline{N}_{out} = \frac{\overline{N}_{in}}{Q} = \frac{\overline{N}_{max}}{Q \cdot D^{m/\nu}} \quad (48)$$

and also improves the network density-outage probability tradeoff (i.e. (23), (24)),

$$\overline{N}_{in} = \int_{V(r(D))} \rho dV \leq Q\epsilon \quad (49)$$

$$\rho \leq Qc_m^{-1}\epsilon \left(P_{\max}/P_t a_\nu\right)^{m/\nu} \quad (50)$$

i.e. the network density $\rho$ can be increased by a factor of $Q$ at the same performance compared to the case of no filtering. It should be noted that $Q$ is similar to an antenna gain (see [19], [20] for detailed discussion of antenna-related concepts). In particular, using highly-directional antennas results in high $Q$ [15]-[17] and thus the network density can be increased by a large factor $Q$, as expected intuitively. A detailed analysis of $Q$ for many popular antenna types can be found in [15]-[17].

Comparing the effect of linear filtering in (48) to that of complete cancellation of $(k-1)$ nearest interferers in (19) and to partial cancellation in (25), it is clear that the complete cancellation (or partial cancellation when the cancellation level is sufficient, i.e. as in (29)) is the most superior technique (scale exponentially with $k$, resulting in significant decrease in the outage probability), and that the linear filtering and partial cancellation are somewhat similar in their effect on the outage probability (scale polynomially with $\alpha$ and $Q$).

## VII. OUTAGE CAPACITY

In this section, the outage capacity is evaluated based on the outage probability expressions above and using the method in [21]. For a given realization of interferers' location and assuming Gaussian signalling, the instantaneous link capacity of a given user can be expressed as $C = \ln(1 + \mathsf{SINR})$ in [nat/s/Hz], where $\mathsf{SINR} = P_s/(P_0 + P_I)$ is the signal to interference plus noise ratio, and $P_s, P_I$ are the signal and interference power. In the interference-dominated scenario, $P_0 + P_I \approx P_I$ so that $\mathsf{SINR} \approx \gamma/d$, where $\gamma = P_s/P_0$, $d = P_I/P_0$ are the SNR and INR. We assume that the SNR is fixed and the INR follows one of the distributions given above.

In terms of the capacity, the outage probability is the probability that the link is not able to support a given rate $R$, $\mathcal{P}_{out} = \Pr\{C < R\}$, and the outage capacity $C_\epsilon$ is the largest rate for which the outage probability does not exceed $\epsilon$ [21], which can be determined from $\mathcal{P}_{out} = \Pr\{C < C_\epsilon\} = \epsilon$ in combination with (11), (12) as

$$C_\epsilon = \ln\left(1 + \frac{\gamma}{D_\epsilon}\right) \quad (51)$$

where $D_\epsilon = F_d^{-1}(1-\epsilon)$ is the outage INR, i.e. the distortion-free INR required to support the outage probability $\epsilon$, and $\gamma/D_\epsilon$ is the signal-to-interference ratio required to support the outage capacity $C_\epsilon$. At high and low SIR, this can be approximated as

$$C_\epsilon \approx \ln\gamma - \ln D_\epsilon, \ \gamma \gg D_\epsilon \quad \text{(high SIR)} \quad (52)$$
$$\approx \frac{\gamma}{D_\epsilon}, \ \gamma \ll D_\epsilon \quad \text{(low SIR)} \quad (53)$$

Note that $\ln\gamma$ and $\gamma$ are the AWGN channel capacity at high and low SNR, and $\ln D_\epsilon$, $D_\epsilon$ represent the capacity loss due to interference, which is additive at high and multiplicative at low SIR.

To see the effect of interference cancelation on the outage capacity, we use (19) to obtain $D_\epsilon \approx \overline{N}_{max}^{\nu/m}/(k!\epsilon)^{\frac{\nu}{mk}}$ so that

$$C_\epsilon \approx \ln\gamma - \frac{\nu}{m}\ln\overline{N}_{max} - \frac{\nu}{mk}\ln(k!\epsilon) \quad \text{(high SIR)}$$
$$\approx \gamma\frac{(k!\epsilon)^{\frac{\nu}{mk}}}{\overline{N}_{max}^{\nu/m}} \quad \text{(low SIR)} \quad (54)$$

Thus, while the outage capacity loss is additive and scales as $\frac{\nu}{mk}\ln(k!\epsilon)$ at high SIR, i.e. roughly linear in $1/k$, it is multiplicative and scales as $(k!\epsilon)^{\frac{\nu}{mk}}$, i.e. exponentially, at low SIR. From this, we conclude that the effect of interference is much more dramatic at low SIR. In this respect, the effect of interference is similar to the effect of fading of the required signal (see [21] for a discussion of the latter).

Using (51)-(53) in combination with the results in Sections IV-VI, the impact of other types of interference cancelation, either alone or in combination with fading, can also be analyzed.

## VIII. CONCLUSION

A model of interference in wireless networks with Poisson spatial distribution of the nodes is considered, which includes the average propagation path loss and also different types of fading. Since the total interference power is dominated by the nearest interferer, the latter is used to define the outage probability. This simplifies the analysis significantly, results in compact, closed-form characterisation of the outage probability, including the case where some interferers are cancelled, either completely or partially, and allows to compare different cancellation strategies and to find the required level of cancellation. The effect of fading is characterized for a broad class of distributions, including all popular fading models and in combination with the effect of interference cancellation. The effect of linear filtering at the receiver (e.g. by directional antennas) is quantified via a new statistical filter gain, and also compared to that of complete/partial cancellation of nearest interferers. These results allow one to express the tradeoff between the node density and the outage probability in an explicit, closed form for a number of scenarios.

Our main findings in terms of the node density - outage probability tradeoff at the low outage region can be summarized as follows:

- for given maximum acceptable outage probability $\epsilon$, $\mathcal{P}_{out} \leq \epsilon$, the upper bound on the node density scales as $\epsilon$ without interference cancellation (see (18));
- when $(k-1)$ strongest interferers are cancelled completely or near completely (see (28), (29)), the upper bound scales as $\epsilon^{1/k}$, i.e. much higher node density can be tolerated (see (24));
- when strongest interferers are partially cancelled by the level independent of the INR, the upper bound still scales as $\epsilon$, with a fixed improvement due to interference cancellation (see (25), (26));
- with linear filtering, the upper bound scales as $\epsilon$, with a fixed improvement due to the filtering (see (50));
- when fading is present, the scaling above still holds (with an additional fixed multiplicative constant, which depends on fading distribution - see (41)-(43)).

Thus, the main conclusion here is that complete or near-complete cancellation of nearest interferers is essential to go from $\epsilon$ to $\epsilon^{1/k}$ scaling.

## IX. Acknowledgement

We would like to thank S. Primak and M. Haenggi for stimulating discussions, and the reviewers for numerous constructive comments.

## X. Appendix 1

*Proof of the Theorem 1*: we need the following lemma (Lemma 4.4.2 in [14]),

*Lemma 1:* Let $X$ be a positive random variable with a regularly varying tail, i.e. there is a number $b > 0$ such that $\forall a > 1$,

$$\lim_{x \to \infty} \frac{\Pr\{X > a \cdot x\}}{\Pr\{X > x\}} = a^{-b} \quad (55)$$

and let the tail of $X$ to dominate the tail of another positive random variable $Y$, i.e.

$$\lim_{x \to \infty} \frac{\Pr\{Y > x\}}{\Pr\{X > x\}} = 0 \quad (56)$$

Then

$$\lim_{x \to \infty} \frac{\Pr\{X + Y > x\}}{\Pr\{X > x\}} = 1. \quad (57)$$

∎

It is straightforward to verify that the tail of $P_{a1}$ dominates the tail of $P_{a2}$ and also the tail of $(N-1)P_{a2}$ for any finite $N \geq 2$ (i.e. that (56) holds with $X = P_{a1}$, and $Y = P_{a2}$ or $Y = (N-1)P_{a2}$) and, thus,

$$\lim_{x \to \infty} \frac{\Pr\{P_{a1} + P_{a2} > x\}}{\Pr\{P_{a1} > x\}} =$$
$$= \lim_{x \to \infty} \frac{\Pr\{P_{a1} + (N-1)P_{a2} > x\}}{\Pr\{P_{a1} > x\}} = 1 \quad (58)$$

Combining this with the following bounds,

$$\Pr\{P_{a1} + P_{a2} > x\} \leq \quad (59)$$
$$\Pr\left\{\sum_i P_{ai} > x\right\} \leq \Pr\{P_{a1} + (N-1)P_{a2} > x\}$$

and noting that $N$ is finite with probability 1 when the average number of nodes is finite, one obtains (13). While (59) formally does not hold when $N = 0$ or 1, there is nothing to prove in such cases as the maximum and total interference powers coincide. $Q.E.D.$

As a side remark, we note that Theorem 1 holds for a broad class of scenarios where the distribution of interfering signal powers can be represented via the functions of regular variations, and not only for the scenario we consider here. In particular, this includes signals subject to Rayleigh and log-normal fading.

## XI. Appendix 2

*Proof of (33):* Consider the outage probability in (32)

$$\begin{aligned} \mathcal{P}_{out} &= \int_0^\infty f_{gs}(g)\overline{F}_d(D/g)\,dg \quad (60) \\ &= \underbrace{\int_0^{D^\varepsilon} f_{gs}(g)\overline{F}_d(D/g)\,dg}_{I_1} + \underbrace{\int_{D^\varepsilon}^\infty f_{gs}(g)\overline{F}_d(D/g)\,dg}_{I_2} \end{aligned}$$

where $0 < \varepsilon < 1$, and note (using (16)) that, when $D \to \infty$,

$$\begin{aligned} I_1 &\approx \frac{\overline{N}_{\max}}{D^{m/\nu}} \int_0^{D^\varepsilon} g^{m/\nu} f_{gs}(g)dg \\ &\approx \frac{\overline{N}_{\max}}{D^{m/\nu}} \int_0^\infty g^{m/\nu} f_{gs}(g)dg \\ &= \Gamma(m/\nu + 1)\frac{\overline{N}_{\max}}{D^{m/\nu}} \quad (61) \end{aligned}$$

On the other hand,

$$\begin{aligned} I_2 &= \int_{D^\varepsilon}^\infty f_{gs}(g)\overline{F}_d(D/g)\,dg \\ &\leq \int_{D^\varepsilon}^\infty f_{gs}(g)dg = e^{-D^\varepsilon} \ll I_1 \quad (62) \end{aligned}$$

and, thus, $I_2$ can be neglected and (33) follows. $Q.E.D.$

Not only this proof gives a compact approximation for the outage probability, but also tells us why this approximation holds: since the tail of the fading distribution decays much faster than that of $P_{a1}$ (compare (61) to (62)), the dominant contribution to outage events is coming from the nearest interferer that is not in deep fade.

It is clear that the same argument also holds when $(k-1)$ nearest interferers are cancelled, when the fading is log-normal or combined (log-normal+Rayleigh), or when the fading process is from the broad class in (40) (the latter three require for a slight modification of the upper bound in (62), which is left as an exercise to the reader).

**Vladimir Mordachev** was born in Vitebsk, Belarus. He received the Ph.D. degree (1984), an academic rank of Senior Scientist (1985), and the M.S. degree with honors (1974) in Radio Engineering from the Minsk Radio Engineering Institute, Minsk, Belarus. His research interests include spectrum management, wireless communications and networks, electromagnetic compatibility and interference, wireless network planning, computer-aided analysis and design, cellular networks system ecology, RF systems modeling and simulation. He is extensively involved in consulting to wireless network operators, industry and the local government. V. Mordachev is a head of the Electromagnetic Compatibility Laboratory at the Belorussian State University of Informatics and Radioelectronics.

**Sergey Loyka** (M'96–SM'04) was born in Minsk, Belarus. He received the Ph.D. degree in Radio Engineering from the Belorussian State University of Informatics and Radioelectronics (BSUIR), Minsk, Belarus in 1995 and the M.S. degree with honors from Minsk Radioengineering Institute, Minsk, Belarus in 1992. Since 2001 he has been a faculty member at the School of Information Technology and Engineering, University of Ottawa, Canada. Prior to that, he was a research fellow in the Laboratory of Communications and Integrated Microelectronics (LACIME) of Ecole de Technologie Superieure, Montreal, Canada; a senior scientist at the Electromagnetic Compatibility Laboratory of BSUIR, Belarus; an invited scientist at the Laboratory of Electromagnetism and Acoustic (LEMA), Swiss Federal Institute of Technology, Lausanne, Switzerland. His research areas include wireless communications and networks, MIMO systems and smart antennas, RF system modeling and simulation, and electromagnetic compatibility, in which he has published extensively. Dr. Loyka is a technical program committee member of several IEEE conferences and a reviewer for numerous IEEE periodicals and conferences. He received a number of awards from the URSI, the IEEE, the Swiss, Belarus and former USSR governments, and the Soros Foundation.